\documentclass[letterpaper,12point,epsfig]{article}
\usepackage{epsfig}

\begin{document}
\begin{center}
\huge{The Fermi Statistics of a Weakly Excited Column of Granular}
\huge{Particles in a Vibrating Bed}

\end{center}
\vspace{.2 cm}
\begin{center}
\large{Holly Kokstein}
\vspace {.2 cm}

\large{Paul V. Quinn Sr.}

\vspace{.2 cm}

\small{Department of Physical Sciences, Kutztown University, Kutztown, 
Pennsylvania 19530} 
\end{center}

\begin{abstract}
A one dimensional experiment in granular dynamics is carried out to test
the thermodynamic theory of weakly excited granular systems [Hayakawa
and Hong, Phys. Rev. Lett. 78, 2764(1997)] where granular particles are
treated as spinless Fermions. The density profile is measured and then fit
to the Fermi distribution function, from which the global temperature of
the system, $T$, is determined. Then the center of mass, $< z(T) >$, and
its fluctuations, $< \Delta z(T)^2 >$, are measured and plotted as functions of $T$.
The Fermi function fits the density profile fairly well, with the value of $T$
being reasonably close to the predicted value. The scaling behavior of $< z(T) >$
and $< \Delta z(T) ^2 >$ is in excellent agreement with the theory.
\end{abstract}



\section{Introduction}
	This paper is an experimental test of the theory presented and tested in preceeding papers [1] and [2].  
The thermodynamic theory of Hayakawa and Hong(HH) presented in [2] was tested in [1] with extensive Molecular 
Dynamics simulations. The purpose of this paper is to test the theory of HH experimentally for a 
one dimensional vibrating granular system. A one dimensional system is peculiar in the sense that 
ran­domness associated with collisions is suppressed in contrast with what happens in higher
dimensions. Nevertheless, the Fermi statistics arising from hard core repulsion still apply to
this simple one dimensional system. As in [1], we determine the config­urational statistics of a 
one dimensional vibrating granular system by properly taking the ensemble average of the steady state. 
Then we measure physical quantities of the system and com­pare them to those predicted by HH. First, we 
measure the density profile and determine the dimensionless Fermi temperature,

$$T = \frac{T_f}{mgD},$$

\noindent where $T_f$ is the Fermi temperature, by fitting
the density profile to the Fermi function. Second, we compare the measured Fermi temperature 
$T$ to those predicted by the theory in [1] and [2]. Third, we measure the center of mass $<z(T)>$ and the 
fluctuations of the center of mass $<\Delta z(T)>$ for the vibrating bed and test the scaling 
predictions of [2].  We will first summarize the theory presented by HH.

\section{Background of Fermi Statistics and Thermodynamic Theory of Granular Materials}

The system being studied here is a dense, dissipative, nonequilibrium, granular system, where the mean free 
path of the grains is of the order of a few particle diameters. Hence, each particle may be considered to be 
effectively confined in a cage as in the free volume theory of a dense liquid [3]. In such a case, an 
observation has been made in [4] that the basic granular state is not a gas, but a solid or crystal, and 
thus, the effective thermodynamic theory based on the free energy argument may be more appropriate than the 
kinetic theory in studying this state. In such a case, the configurational statistics of the steady state may be 
determined by the variational method as the most probable or minimum free energy state.

To be more specific, consider the excitation of disordered granular mate­rials confined in a box with vibrations 
of the bottom plate. The vibrations will inject energy into the system which cause the ground state to become 
unstable, and a newly excited state will emerge with an expanded volume. The time averaged configurational 
statistics of this new excited state have un­dergone structural distortions. However, the degree of distortions 
from the ground state may be small for a weakly excited state, possibly justifying the use of an effective 
thermodynamic theory based on the variational principle. Such a thermodynamic approach may be further justified 
by the following two experiments conducted previously.

	\bf{1. Weakly or moderately excited regime:} \normalfont Clement and Rajchenbach(CR) [4] have performed an 
experiment with the vibrational strength, $\Gamma$, of the order one for a two dimensional vibrating bed, using 
inclined side walls to suppress convections. Here, $\Gamma = A \omega^2 / g$ with $A$ and $\omega$, the amplitude 
and frequency of the vibrating plane, and $g$, the gravitational constant. CR have found that the 
ensemble-averaged density profile as a function of height from the bottom layer obeys a universal function that 
is independent of the phase of oscillations of the vibrating plate. Namely, it is independent of the kinetics 
imposed on the system. One conceptually important point here is that the reference point of the density profile 
is not the bottom plate, but the bottom layer, which of course is fluidized.

	\bf{2. Highly excited regime:} \normalfont Warr and Hansen(WH) [5] have performed an experiment on highly 
agitated, vertically vibrating beds of $\Gamma \approx 30 - 50$ using steel balls with a small coefficient of 
restitution. They have found that the collective behaviors of this vibrated granular medium \it{in a station­ary 
nonequilibrium state} \normalfont exhibits strong similarities to those of an atomistic fluid in \it{thermal 
equilibrium} \normalfont at the corresponding particle packing fraction, in particular, in the two-point 
correlation function [6,7].

The results of both experiments indicate that for both moderate or highly excited systems, one-to-one 
correspondence seems to exist between \it{configu­rational} \normalfont statistics of the nonequilibrium 
stationary state and the equilibrium thermal state. In fact, this is not so surprising considering that upon 
vibra­tion, the granular materials expand and increase the volume of the system. In turn, this increase 
corresponds to a rise in the potential energy after the configurational average is appropriately taken. Then the 
problem reduces to the packing problem, and the temperature-like variable, $T$, can be as­sociated to the vibrating 
bed. The existence of distinctive configurational statistics in the density profile of CR (and also in WH in a 
special case) appears to be fairly convincing evidence that kinetic aspects of the excited granular materials may 
be separated out from the statistical configurations. These observations are the basis of the thermodynamic 
theory proposed in [2].  Note that the Fermi statistics is essentially the macroscopic manifesta­tion of the 
classical excluded volume effect and the anisotropy which causes the ordering of potential energy by gravity. The 
top surface of the granules plays the role of a Fermi surface, and the thin boundary layers that appear near the 
top layer upon excitation play the role of excited electrons of the Fermi gas in metals.

\section {Thermodynamic Theory of Weakly Excited Granular Systems}

\bf{1. Fermi temperature:} \normalfont In [2], the vibrating system was viewed from two different points of 
view.  One may view it as a mechanical system, in which case the expansion is due to excitation induced by 
mechanical vibrations of strength $\Gamma = A \omega^2 / g$. In this case, the expansion is purely due to 
kinetics. In an attempt to develop a thermodynamic theory, such a system was also viewed [2] as a thermal system 
in contact with a heat reservoir. Therefore a global temperature $T$ was associated with the granular system. In 
this case, the expansion is purely a thermal expansion. By equating the thermal expansion, $\Delta h$, defined as 
the increase in the center of mass, and the kinetic expansion, $g H_o(\Gamma) / \omega^2$, where $H_o(\Gamma)$ is 
the dimensionless jump height of a single ball [2,8] on the vibrating plate(see Eq.(l) in [2]), a closure in the 
thermodynamic theory of powders was obtained in [2]. Since the density decrease above the Fermi surface is not 
sharp, but smooth, one may replace $g H_o(\Gamma) / \omega^2$ with $h_o(\Gamma)/ \alpha$, where $h_o(\Gamma)$ is 
the maximum jump height of a single ball at the Fermi surface ( or vibrating plate) determined by MD simulations. 
The factor $\alpha$ was introduced to both expressions to incorporate (i) the smooth decrease in the density 
profile near the Fermi surface, and (ii) the suppression in the jump height due to dissipation. By equating the 
kinetic expansion and the thermal expansion, one obtains the following explicit relationships between the 
temperature $T$ and the control parameters:

$$ \frac {T}{mg} = \frac{1}{\pi} \sqrt{ \frac{6 D [g H_o(\Gamma) / \omega^2 ]}{\alpha}}, \eqno(1a)$$
$$ \frac {T}{mg} = \frac{1}{\pi} \sqrt{ \frac{6 D [h_o(\Gamma)}{\alpha}}. \eqno(1b) $$

Note that when a single particle is on a vibrating plate, the energy from the bottom wall is transferred to the 
particle through direct contact. In the case of many particles, the supplied energy at the vibrating plate must 
first travel through other particles locked in their respective lattice states before reaching those particles in 
the fluidized layer.  With this in mind, $\alpha$ can be thought of as a dissipation constant. In [1], $\alpha$ 
was empirically determined to have a value of $\alpha = 64/5$.  This value of $\alpha$ fit the data in [1], 
regardless of the type of vibration used. 

\bf{2. The center of mass and its fluctuations:}\normalfont Since the density profile is given by the Fermi 
function, it is straightfoward to compute the center of mass, $< z(T) >$, and its fluctuations, 
$<(\Delta z)^2 >$.  This leads to the following equations: 

$$\Delta z(T) = z(T) - z(0) = \frac{D \mu_o \pi^2}{6}(\frac{T}{mgD\mu_o})^2, \eqno(2a)$$
$$<(\Delta z)^2> = <(z(T)-< z >)^2> = \frac{<(\Delta h)^2>}{\mu_o^2} = \frac{\pi^2}{3}
(\frac{T}{mgD})^3\frac{D^2}{\mu_o^2}. \eqno(2b)$$

\noindent Note that the total expansion, $\Delta h(T) \equiv \mu_o \Delta z$, and its fluctuations 
$<(\Delta h)^2> / D^2 =<\mu_o(\Delta z)^2 / D^2 >$ are only a function of the dimensionless Fermi tem­perature 
$T_f = T /mgD$ as expected. Further, note that Eq.(2b) is an indirect confirmation that the specific heat is 
linear in $T$ as it is for the non-interacting Fermi gas.

\section {The One Dimensional Experiment}

We can now explain the experimental setup in some detail. The one dimensional column of
particles used in this experiment was $30$ plastic beads. The beads, having holes through
their centers, were strung through a thin piece of copper wire. The wire was stretched
extremely tight between two horizontal rods clamped to a ring stand. Each of the beads could freely 
move up and down the copper wire. The beads were all close to the same size, with an average diameter of 
$D=4.24$ $mm$. Differences in the diameters of the beads were not noticeable to the naked eye. The bottom of the 
copper wire ran through a plastic plate connected directly to the mechanical vibrator, which is very similar to 
an audio speaker. One of the plastic beads was glued to the plastic plate, acting as the bottom wall when the 
vibration was turned on. A schematic diagram of the set up is shown in Fig.1.  The mechanical vibrator was then 
connected to a function generator, which allowed one to control the type, amplitude, and frequency of the 
mechanical vibration. For this experiment, the only type of vibration used was a sine wave, $A sin(\omega t)$, 
where $A$ is the amplitude and $\omega = 2 \pi f$ is the angular frequency. The frequency of the vibration and 
the voltage which set the amplitude were both controlled with the function generator. The relationship between 
the voltage and the vibration amplitude is nonlinear and quite com­plex. Hence, instead of deriving the 
mathematical formula, for a given applied voltage, we directly measured the corresponding amplitude of the 
vibration of the plate. We find that voltage with a range of $1.00 V$ to $10.00 V$ corresponds to an amplitude of 
$0.495$ $mm$ to $1.890$ $mm$. In this experment, the frequency $f$ was kept at $f=40 Hz$ while the voltage was 
varied by $0.5 V$ increments from $1.00 - 10.00 V$. The amplitudes, $A$, and the corresponding vibrational strength 
$(\Gamma = A \omega^2 / g)$ with $g$ being the gravitational acceleration are listed in Table 1. This set of 
control parameters satisfies the criterion of the weakly excited granular system used to test the theory of 
HH, namely that

$$R = \frac{\mu D}{\Gamma A}, \eqno(3)$$

\noindent with $\mu$ being the number of layers. To analyze the system, pictures were taken with
a high speed digital camera while the beads were subjected to vibration. The camera was placed on a stand in 
front of the beads, while pictures were taken of the system in motion.  This is schematically illustrated in 
Fig.2. A set of $10$ pictures was taken for each value of $\Gamma$ used in this experiment, as well as a picture 
of the beads at rest in the begining and end of each run. Then, all the images were analyzed to obtain the 
position of the center of each bead for each image.  Once the individual bead coordinates were known, they were 
fed into a computer program to calculate the avaerage density profile of the system, the average center of mass, 
and the fluctuations of the center of mass for each voltage setting.  

To find the density profile, equally sized boxes with a height equal to the average diameter of a bead were 
constructed so that each box essentially contained, one bead when the system was at rest in the ground state. 
Note that the zero reference point in all our measurements is not the vibrating plate directly, but rather a 
single particle glued to the bottom.  This was done to ensure that all collisions were the same for all 
particles, including the bottom wall collision.  

Before proceeding, we want to clarify two points regarding the density measurement. First, the maximum square 
packed density for a one dimensional system of spheres is calculated to be $\pi/6$.  With our experiment, we used 
pictures of the system to analyze the density.  Since the pictures are two dimensional, we can treat our column 
of spherical beads as column of disks confined to one dimensional motion.  Using this analysis, the maximum 
square packed density for our column of disks is calculated to be $\pi/4$.  However, on the density plots 
presented in this paper, the density exceeded this value for certain data points. This is because box sizes 
were assumed to all have a length and width equal to the average bead radius. In reality, however, this was not 
true, leading at times to a density greater than $\pi/4$. Second, since the bottom layer is the reference 
point, one may expect that the density of the first data point should remain the same.  However, we noticed that 
the values of the density of the bottom changed slightly in our data. This was due to the erosion of the balls 
that occurred due to the repeated motion on the copper wire. This erosion of the center hole caused the 
configuration of the bottom few balls to become slightly more dense as more trials were completed.  

For this experiment, $50$ boxes were chosen because the motion of the column of beads never reached that height, 
even for the largest voltage.  Therefore, every bead was included in the calculations of all the density 
profiles. Since this is a one dimensional experiment, each bead was considered to be a circle instead of a 
sphere. Hence, the maximum possible area of a bead in a box, $a_o$, is $\pi r^2$ where 

$$r =< D > /2 = \Sigma_i D_i / 2N$$ 

\noindent is the average radius of a bead.  $D_i$ is the diameter of bead $i$, and $N$ is the total number of 
beads. In the ground state, when the system is at rest, each box up to the Fermi surface contains one bead with 
an area of $a_o$.  However, when the beads are subjected to vibration, the system expands, and the time averaged 
position of each grain rises. The area of the grains in each box is computed from the pictures, which is denoted 
as $a$. Then, the density of each box, having an approximate value between $0$ and $1$, was computed using

$$\rho = \frac{a}{a_o}. \eqno(4)$$

\noindent This process was done for all $50$ boxes and in all $10$ pictures for each voltage used. An average 
density for each box was then found by summing up the $10$ densities per box and dividing by $10$. The density 
profile for a particular voltage was obtained by plotting the average density of each box against the position of 
the center of that particular box.  The center of the first box is chosen as the origin.  The profile is then 
fit with the Fermi profile, and a global temperature, $T$, is obtained. The fitting of the density profile was 
done with a non-linear least squares program.  Computer programs were used to obtain the average center of mass 
and its fluctuations. The center of mass is computed using the following expression:

$$z = \frac{\Sigma_i z_i m_i}{\Sigma_i m_i}, \eqno(5)$$

\noindent where $z_i$ is the vertical position in centimeters of the bead $i$ and $m_i$ is the mass in grams
of the bead $i$. These summations were carried out over all $30$ beads. This formula was used to get the center 
of mass for each picture, and then the average was taken over all $10$ pictures to obtain the average center of 
mass, $< z >$. 

The average fluctuations of the center of mass are computed using the standard definition:

$$<(\Delta z)^2> = \frac{[\Sigma_j (z(j) - < z >)]^2}{10}, \eqno(6)$$

\noindent where $z(j)$ is the center of mass of an individual picture, and $< z >$ is the average center of 
mass found previously for a particular value of voltage. In this case, the summation is carried out over $10$, the 
number of pictures taken per run. Both the average center of mass and the fluctuations were computed for each 
voltage setting.

\section{Data and Results}

\bf{Density Profile:} \normalfont This experiment was carried out for $30$ particles using a sine wave vibration 
ranging from $1.00 -  10.00 V$ and varying by increments of $0.5 V$ while being held at a constant frequency of 
$f = 40 Hz$. The change in $V$ is synonymous with a change in the vibrational strength, $\Gamma$. The density 
profile, $\rho$, as a function of the vertical position, $z$, for the different voltages was fit with the Fermi 
profile as shown in Fig.(3). Note that for an electron gas, the Fermi density profile is given by

$$\rho(\epsilon_i) = \frac{1}{1 + exp[(\epsilon_i - \overline{\mu})/T_{f}]}, \eqno(7)$$

\noindent where $\epsilon_i$ is the energy level of electron $i$, $\overline{\mu}$ is the 
Fermi energy as a function of $T$, and $T_f$ is the Fermi temperature. As discussed in [1], for granular systems, 
$\overline{\mu}(T)$ is independent of the temperature because the density of the state is constant. We denote 
this constant as $\mu_o \equiv \overline{\mu}(T = 0)$, which is the initial number of layers. Note that the 
dimensionless Fermi energy for our granular system at $T=0$ is the initial number of layers, while the Fermi 
energy $\overline{\mu}$, is replaced with $mgD\mu(T)$. For the configurational statistics of granular materials 
in a vibrating bed, the energy level is given by the average position of the particle. Thus, the energy level is 
given by the gravitational energy, $\epsilon_i = mgz_i\ equiv mgD\overline{z}$, where $z_i$ is the vertical 
position of particle $i$ and $\overline{z}$ is the dimensionless position measured in units of $D$. Note that 
$\overline{z}$, runs from $0$ to $50$ because the position of the first grain is taken as the origin. The 
relationship between the dimensionless fitting Fermi temperature $T$ in this paper and the Fermi temperature 
$T_f$ in Eq.(2a) is given by $T = T_f/mgD$.  This leads to the following density profile, 

$$\rho(z) = \frac{1}{1 + exp[(\epsilon_i(\overline{z}) - \overline{\mu})/T_f]} 
\equiv \frac{1}{1 + exp[(\overline{z}-\mu)/T} \eqno(8)$$

An example of Eq.(8) fit to our data can be seen in Fig.3, which is for a particular voltage of $10 V$.  Note 
that the Fermi statistics are valid for all of our data according the to condition that $R \gg 1$ as stated in 
Eq.(3).  In our experiment, $R$ changes from $76.3$ to $5.6$ as shown in Table 2.

\bf{Fermi temperature fitting:}\normalfont   When fitting the experimental data with the Fermi func­tion, the 
following two adjustable parameters are used: the Fermi energy, $\mu$, and the global tem­perature $T$ as 
defined previously. The parameter $\mu$ shifts the location of the Fermi energy horizontally, while the
temperature, $T$, controls the curvature around the Fermi energy. As discussed above, the Fermi energy is 
expected to remain constant regardless of the temperature $T$ if the density of states is indeed constant.  
However, the experimental fitting shows that there is some slight variation in $\mu$. In this experiment, $\mu$ 
varies from $29.01$ to $30.57$, a relative increase of about $\Delta \mu / \mu = 0.051$. The scaling behavior of 
the center of mass and its fluctuations don't seem to be affected by the change in $\mu$. The fitting values 
obtained for $\mu$ and $T$ are also listed in Table 2.

 When the temperature, $T$, obtained by the Fermi fitting is compared to the to the theoretical prediction from 
Eq.(1a), it seems to match fairly well. As discussed in [1], a best fit value of $\alpha = 64/5$ was used to 
obtain the theoretical temperature. Table 3 shows the values of the fitting temperature $T$, and two theoretical 
temperatures $T_{theory}1$ and $T_{theory}2$.  Both are obtained by using Eq.(1a).  The difference in the two 
predicted values is because of differences in the dissipation constant $\alpha$.  For $T_{theory}1$, the value of 
$\alpha=64/5$ was used, which was the value found empirically using MD simulations in [1].  This value of 
$\alpha$ provided predicted values all within $30 \%$ of the measured $T$.  However, it is not necessarily 
correct to assume that energy will be dissipated through our experimental grains exactly the same way as it was 
in the MD simulations of [1].  Depending on how the constants in the MD program are set, the particle reactions 
during a collision could be quite different then that which occured in our experiment.  Hence, $T_{theory}2$ was 
obtained with a best fit value of $\alpha=8$, determined empirically from our data.  With this new value of 
$\alpha$, most of the predicted values are well within $10\%$ of what we measured. 

\bf{The Center of Mass:}\normalfont   The relative increase in the center of mass of the one dimensional
column is denoted as $< \Delta z(T) >$, which is the difference in the actual position of the average center
of mass, $ <z(T)>$, and that of the ground state $<z(T=0)>$.  Using our definitions for undimensional 
variables, Eq.(2a) becomes 

$$<\Delta z(T)> = \frac{\pi^2 D}{6 \mu} T^2. \eqno(9)$$

\noindent Note the appearance of the factor $D$ in Eq(9). This is because, in Eq.(2a), the temperature $T$ has 
dimensions of length, while here it is dimenionless. The center of mass is plotted in Fig.4 as a function of $T^2$ 
for different values of $\Gamma$. By using the solid line as a guide for the eye, one can see that 
the graph seems to confirm the scaling predictions of the Fermi statistics, $\Delta z \propto T^2$.  There is, 
however, a discrepency in the proportionality constant $C$. Using the average value of $\mu$ calculated from 
Table 2, theory predicts that one should get $C = \pi^2D/6\mu \approx 0.000234$.  The experimental results yield, 
$0.0001$, a difference of about a factor of $2.5$ when compared to the theory. It has been shown in [1] that this 
discrepancy is due to the extreme sensitivity of the center of mass to the Fermi energy $\mu$.  When the density 
of states is independent of the energy, the Fermi energy must remain constant. However, in this one dimensional 
experiment, it changes ever so slightly for each value of $\Gamma$.  This small change does not seem to change 
the Fermi fitting of our density profiles, but it does affect the amplitude of the scaling relationships.

\bf{Fluctuations of the center of mass:} \normalfont The fluctuation of the center of mass, $< (\Delta z)^2 >$,
is given by Eq.(2b), which becomes 

$$<(\Delta z)^2> = \frac{\pi^2}{3}\frac{D^2}{\mu^2}T^3, \eqno(10)$$ 

\noindent when changing to dimensionless variables.  The fluctuations of the center of mass were also measured 
and plotted as a function of $T^3$ in Fig.5. Using the guide line, the graph seems to confirm the validity 
of the Fermi statistics which imply that $<(\Delta z)^2> \propto T^3$. Once again, the proportionality constant 
deviates from the theoretical value of $6.67 \times 10^{-8}$. From the graph, the slope is approximately 
$1.00 \times 10^{-6}$.  Considering that the amplitude of the center of mass is off by a factor of about $2.5$, 
one should expect that the fluctuations would be off as well. This is, again, due to the sensitivity of the Fermi 
integral to $\mu$, where small changes in $\mu$ are greatly magnified in the fluctuations, resulting in the 
deviation from the theoretical constant. A similar trend was found using MD simulations[1]. Another
source of error is due to the fact that the fluctuations in the center of mass are quite large in the vibrating 
column, and all the particles, not just those near the surface, fluctuate in the continuum space of the 
experiment. This is different from the Fermi model which makes all particles below the Fermi surface, inactive. 
So the positions of the grains, on average, may obey the Fermi distribution function well, but the magnitude of 
the fluctuations may not. It is very suprising, though, that the $T^3$ prediction of the Fermi statistics still 
seems to hold.

\section{Conclusion}

We now summarize the main results of this experiment as follows. First, the configu­rational statistics of grains 
in a one dimensional system subject to vibration seem to obey the Fermi statistics of spinless particles for 
weakly excited systems as was predicted[1,2]. Second, the temperatures determined by fitting the Fermi profile 
are fairly close to the theoretical predictions. Third, the scaling relations of the center of mass and its 
fluctuations obey the Fermi statistics, but there are discrepencies in the amplitudes or proportionality 
constants. Note that another source of error in this experiment results from friction. The beads were confined to 
vibration in a single column by the copper wire that ran through their centers. After many trials, filings from 
the inside of the plastic beads were found at the base of the setup. The motion of the beads on the wire eroded 
away some of the hole in the center of each bead. This friction could be responsible for some of the error
in the results. Nevertheless, we find that the essence of the Fermi statistics survives in this
one dimensional experiment.

\newpage 

\Large
\noindent \bf References
\normalsize
\normalfont
\vskip .4 true cm 

\noindent [1] Paul V. Quinn and Daniel C. Hong, \it{Physica A},
\normalfont \bf{274}, \normalfont 572(1999).

\vskip .2 true cm

\noindent [2] H. Hayakawa and D. C. Hong, \it{Phys. Rev. Lett},
\normalfont \bf{78}, \normalfont 2764(1997).

\vskip .2 true cm

\noindent [3] See, eg. T. L. Hill, \it{Statistical Mechanics},
\normalfont Dover, New York, Chapt. 8(1987).

\vskip .2 true cm

\noindent [4] E. Clement and J. Rajchenbach, \it{Europhys. Lett}, 
\normalfont \bf{16}, \normalfont 1333(1991).

\vskip .2 true cm

\noindent [5] S. Warr and J. P. Hansen, \it{Europhys. Lett}, 
\normalfont \bf{36} \normalfont No.8(1996).

\vskip .2 true cm

\noindent [6] J. P. Hansen and I. R. McDonalds, \it{Theory of Simple Liquids}, 
\normalfont Academic Press, London (1986)
\vskip .2 true cm

\noindent [7] G. Ristow, \it{Phys. Rev. Lett.}, 
\normalfont \bf{79}, \normalfont 833(1997).

\vskip .2 true cm

\noindent [8] J. M. Luck and A Mehta, \it {Phys. Rev. E}, 
\normalfont \bf{48} \normalfont 3988(1993).

\newpage

\Large
\noindent \bf Table Captions
\normalsize
\normalfont
\vskip .4 true cm 

\noindent\bf{Table 1:}\hspace{.25 cm}\normalfont The values of the vibrational amplitude $A$, and the vibrational 
strength $\Gamma$ for each corresponding voltage $V$.

\vskip .2 true cm

\noindent\bf{Table 2:}\hspace{.25 cm}\normalfont  A set of parameters obtained in the experiment for various 
voltages. Here, $\mu$, is the dimensionless Fermi energy, $T$ is the dimensionless Fermi temperature, and $R = 
\mu D / \Gamma A$ is the parameter that determines the validity of the Fermi statistics.

\vskip .2 true cm

\noindent\bf{Table 3:}\hspace{.25 cm}\normalfont  The comparison of the experimentally measured Fermi 
Temperature, $T$, and two theoretical Fermi temperatures determined from Eq.(1a).  The first, $T_{theory}1$, is 
determined using the value of $\alpha =64/5$, as was determined in [1].  The second theoretical temperature, 
$T_{theory}2$ is determined by using a best fit $\alpha = 8$ value, determined empirically from the data.

\vskip .2 true cm

\newpage
\pagebreak
\Large
\noindent \bf Figure Captions
\normalsize
\normalfont
\vskip .4 true cm 

\noindent \bf{Figure 1:}\hspace{.25 cm}\normalfont  An illustration of the experimental setup used to vibrate 
a column of beads at a controlled amplitude and frequency.

\vskip .2 true cm

\noindent \bf{Figure 2:}\hspace{.25 cm}\normalfont  An illustration of the experimental setup used to take 
pictures of the vibrating column of beads with a high speed digital camera.

\vskip .2 true cm

\noindent \bf{Figure 3:}\hspace{.25 cm}\normalfont  The density profile of our one dimensional granular system 
vibrating under $10 V$, which is equivalent to $\Gamma = 12.18$.  The profile is fit with the Fermi density 
function, giving values for the undimensionalized variables $\mu = 30.57$ and $T=4.57$.  One can see the profile 
fits extremely well.

\vskip .2 true cm

\noindent \bf{Figure 4:}\hspace{.25 cm}\normalfont Plots of the average center of mass $< z_{avg} >$ as a function 
of $T^2$.  The data is shown with the circles, while dashed line is the best fit line.  The data matches the 
theoretical prediction of a $T^2$ dependence for the average center of mass.  

\vskip .2 true cm

\noindent \bf{Figure 5:}\hspace{.25 cm}\normalfont Plots of the fluctuations of the average center of mass 
$< (\Delta z)^2 >$ as a function of $T^3$.  The data is shown with the circles, while the dashed line is the best 
fit line.  The data follows the theoretical prediction of a $T^3$ dependence for the fluctuations.   

\newpage

\begin{tabular}{|c|c|c|} \hline
Voltage & Amplitude & $\Gamma$ \\ \hline
$1.5 V$ & $0.50$ mm & $3.22$ \\ \hline
$2.0 V$ & $0.55$ mm & $3.54$ \\ \hline
$2.5 V$ & $0.64$ mm & $4.12$ \\ \hline
$3.0 V$ & $0.71$ mm & $4.57$ \\ \hline
$3.5 V$ & $0.76$ mm & $4.89$ \\ \hline
$4.0 V$ & $0.78$ mm & $5.02$ \\ \hline
$4.5 V$ & $0.86$ mm & $5.53$ \\ \hline
$5.0 V$ & $0.92$ mm & $5.92$ \\ \hline
$5.5 V$ & $0.96$ mm & $6.18$ \\ \hline
$6.0 V$ & $1.05$ mm & $6.76$ \\ \hline
$7.0 V$ & $1.20$ mm & $7.72$ \\ \hline
$7.5 V$ & $1.30$ mm & $8.37$ \\ \hline
$8.0 V$ & $1.40$ mm & $9.02$ \\ \hline
$8.5 V$ & $1.56$ mm & $10.05$ \\ \hline
$9.0 V$ & $1.65$ mm & $10.64$ \\ \hline
$9.5 V$ & $1.80$ mm & $11.60$ \\ \hline
$10.0 V$ & $1.89$ mm & $12.18$ \\ \hline
\end{tabular}

\newpage
\pagebreak
\begin{tabular} {|c|c|c|c|c|}\hline
Voltage & $\Gamma$ & $\mu$ & $T$ & $ R = \frac{\mu D}{\Gamma A}$ \\ \hline
$1.50 V$ & $3.22$ & $29.01$ & $1.16$ & $76.3$ \\ \hline
$2.00 V$ & $3.54$ & $29.01$ & $1.75$ & $63.1$ \\ \hline
$2.50 V$ & $4.12$ & $29.25$ & $2.10$ & $50.0$ \\ \hline
$3.00 V$ & $4.57$ & $29.36$ & $2.62$ & $38.3$ \\ \hline
$3.50 V$ & $4.89$ & $29.41$ & $2.71$ & $33.5$ \\ \hline
$4.00 V$ & $5.02$ & $29.43$ & $2.91$ & $31.8$ \\ \hline
$4.50 V$ & $5.53$ & $29.60$ & $3.06$ & $26.3$ \\ \hline
$5.00 V$ & $5.92$ & $29.72$ & $3.12$ & $23.1$ \\ \hline
$5.50 V$ & $6.18$ & $29.72$ & $3.35$ & $21.2$ \\ \hline
$6.00 V$ & $6.76$ & $29.72$ & $3.49$ & $17.7$ \\ \hline
$7.00 V$ & $7.72$ & $30.07$ & $3.90$ & $13.7$ \\ \hline
$7.50 V$ & $8.37$ & $30.19$ & $3.99$ & $11.8$ \\ \hline
$8.00 V$ & $9.02$ & $30.19$ & $4.22$ & $10.1$ \\ \hline
$8.50 V$ & $10.05$ & $30.19$ & $4.31$ & $8.2$ \\ \hline
$9.00 V$ & $10.64$ & $30.31$ & $4.43$ & $7.3$ \\ \hline
$9.50 V$ & $11.60$ & $30.42$ & $4.49$ & $6.2$ \\ \hline
$10.00 V$ & $12.18$ & $30.57$ & $4.57$ & $5.6$ \\ \hline
\end{tabular}

\newpage
\pagebreak
\begin{tabular} {|c|c|c|c|} \hline
Voltage & $T$ & $ T_{theory}1$ & $T_{theory}2$ \\ \hline
$1.50 V$ & $1.16$ & $1.48$ & $1.87$\\ \hline
$2.00 V$ & $1.75$ & $1.62$ & $2.05$\\ \hline
$2.50 V$ & $2.10$ & $1.86$ & $2.35$\\ \hline
$3.00 V$ & $2.62$ & $2.03$ & $2.57$\\ \hline
$3.50 V$ & $2.71$ & $2.14$ & $2.71$\\ \hline
$4.00 V$ & $2.91$ & $2.19$ & $2.77$\\ \hline
$4.50 V$ & $3.06$ & $2.36$ & $2.98$\\ \hline
$5.00 V$ & $3.12$ & $2.48$ & $3.14$\\ \hline
$5.50 V$ & $3.35$ & $2.56$ & $3.24$\\ \hline
$6.00 V$ & $3.49$ & $2.73$ & $3.46$\\ \hline
$7.00 V$ & $3.90$ & $3.00$ & $3.79$\\ \hline
$7.50 V$ & $3.99$ & $3.16$ & $4.00$\\ \hline
$8.00 V$ & $4.22$ & $3.32$ & $4.20$\\ \hline
$8.50 V$ & $4.31$ & $3.55$ & $4.49$\\ \hline
$9.00 V$ & $4.43$ & $3.69$ & $4.66$\\ \hline
$9.50 V$ & $4.49$ & $3.90$ & $4.93$\\ \hline
$10.00 V$ & $4.57$ & $4.00$ & $5.06$\\ \hline
\end{tabular}
\begin{figure}
\begin{center}
\epsfig{file=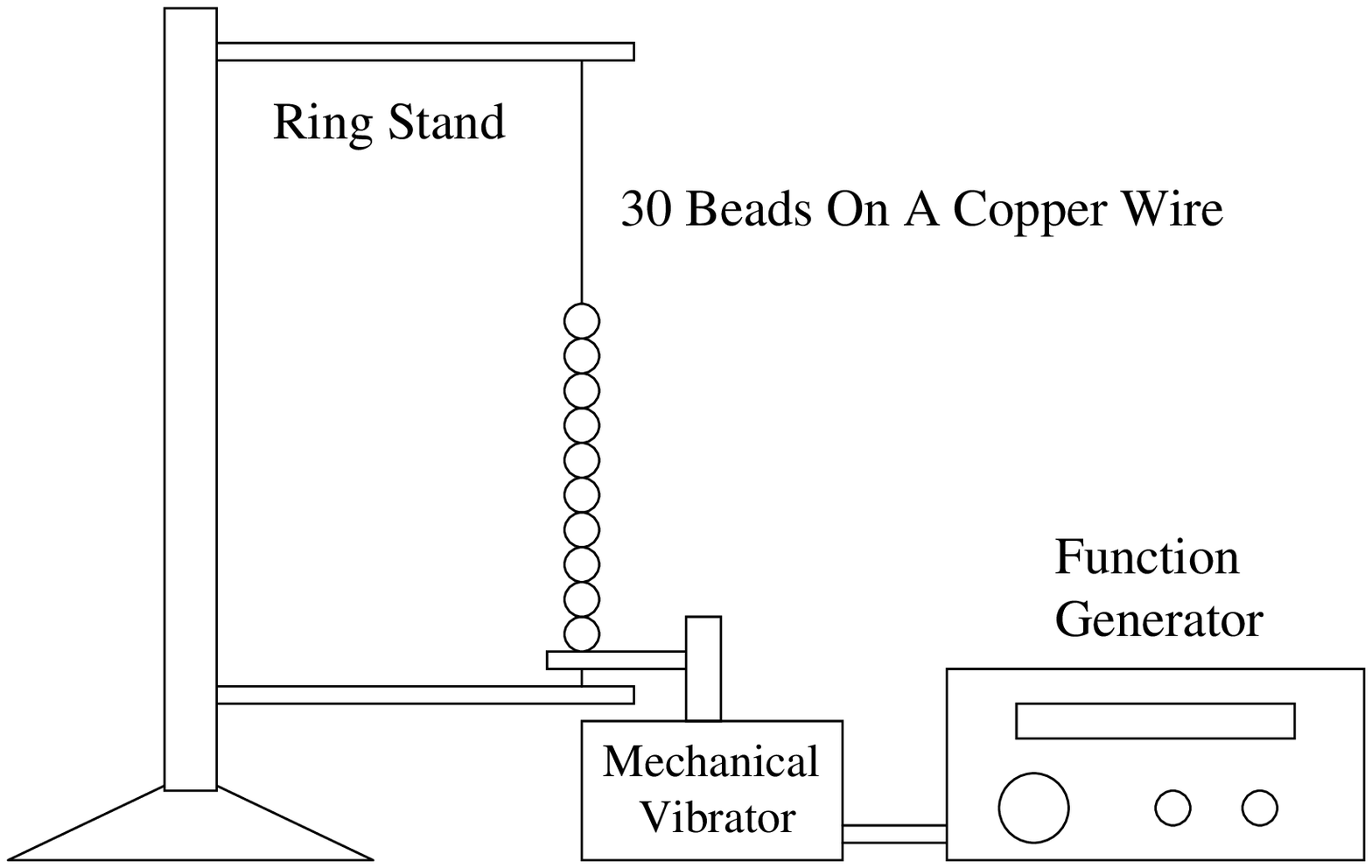,height=15cm}
\end{center}
\end{figure}

\newpage
\pagebreak
\begin{figure}
\begin{center}
\epsfig{file=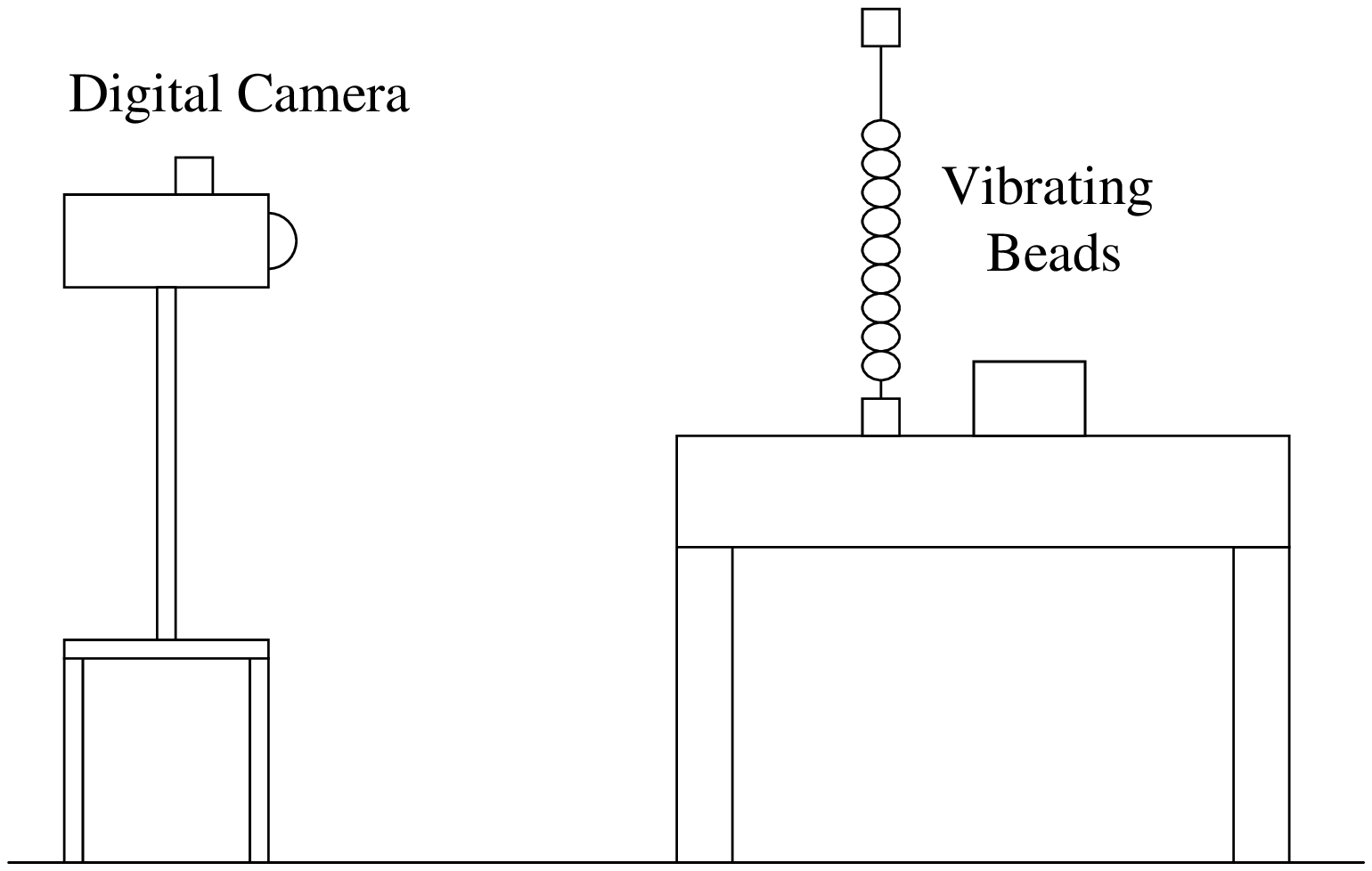,height=15cm}
\end{center}
\end{figure}

\newpage
\pagebreak
\begin{figure}
\begin{center}
\epsfig{file=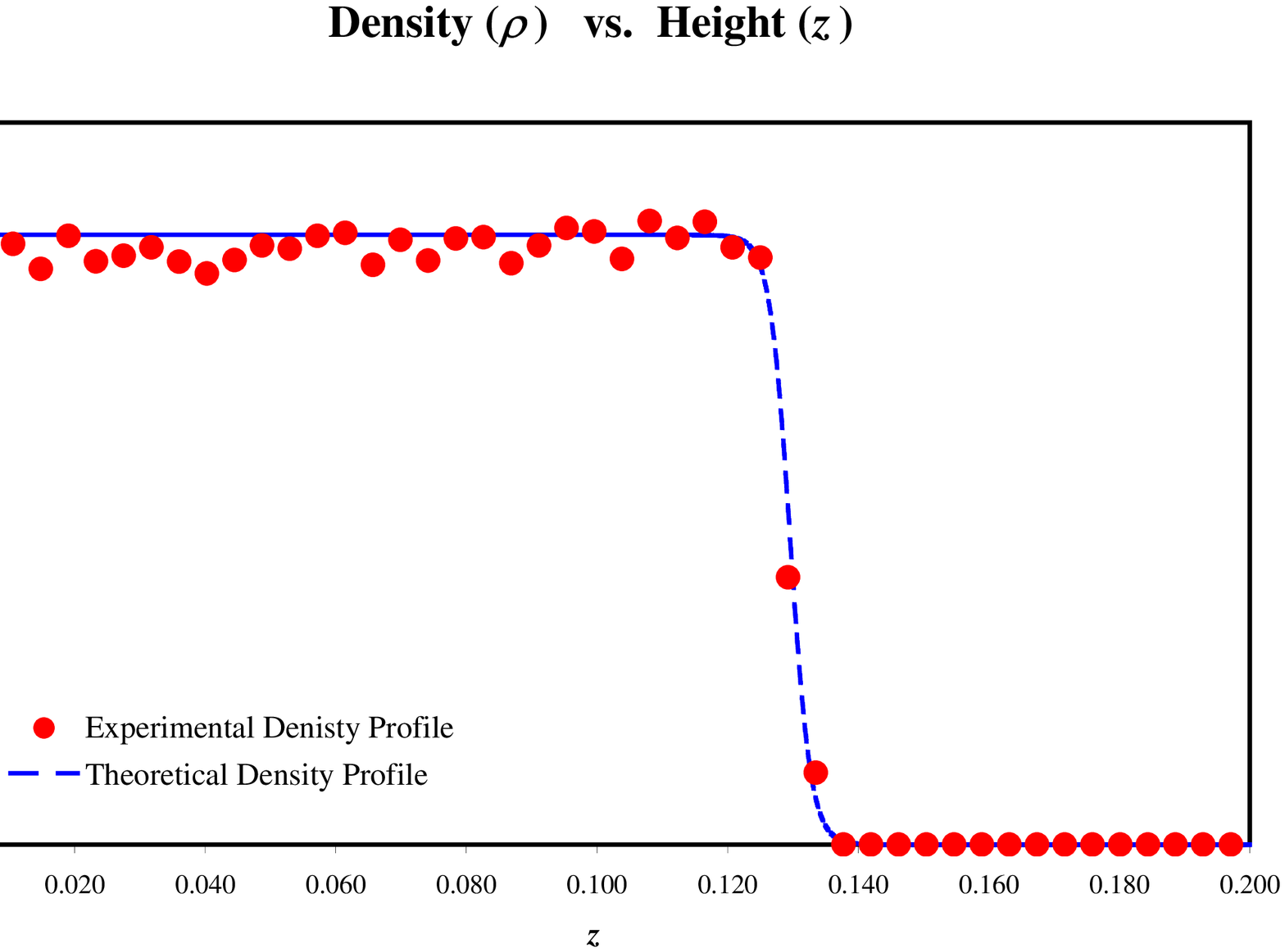,height=15cm}
\end{center}
\end{figure}

\newpage
\pagebreak
\begin{figure}
\begin{center}
\epsfig{file=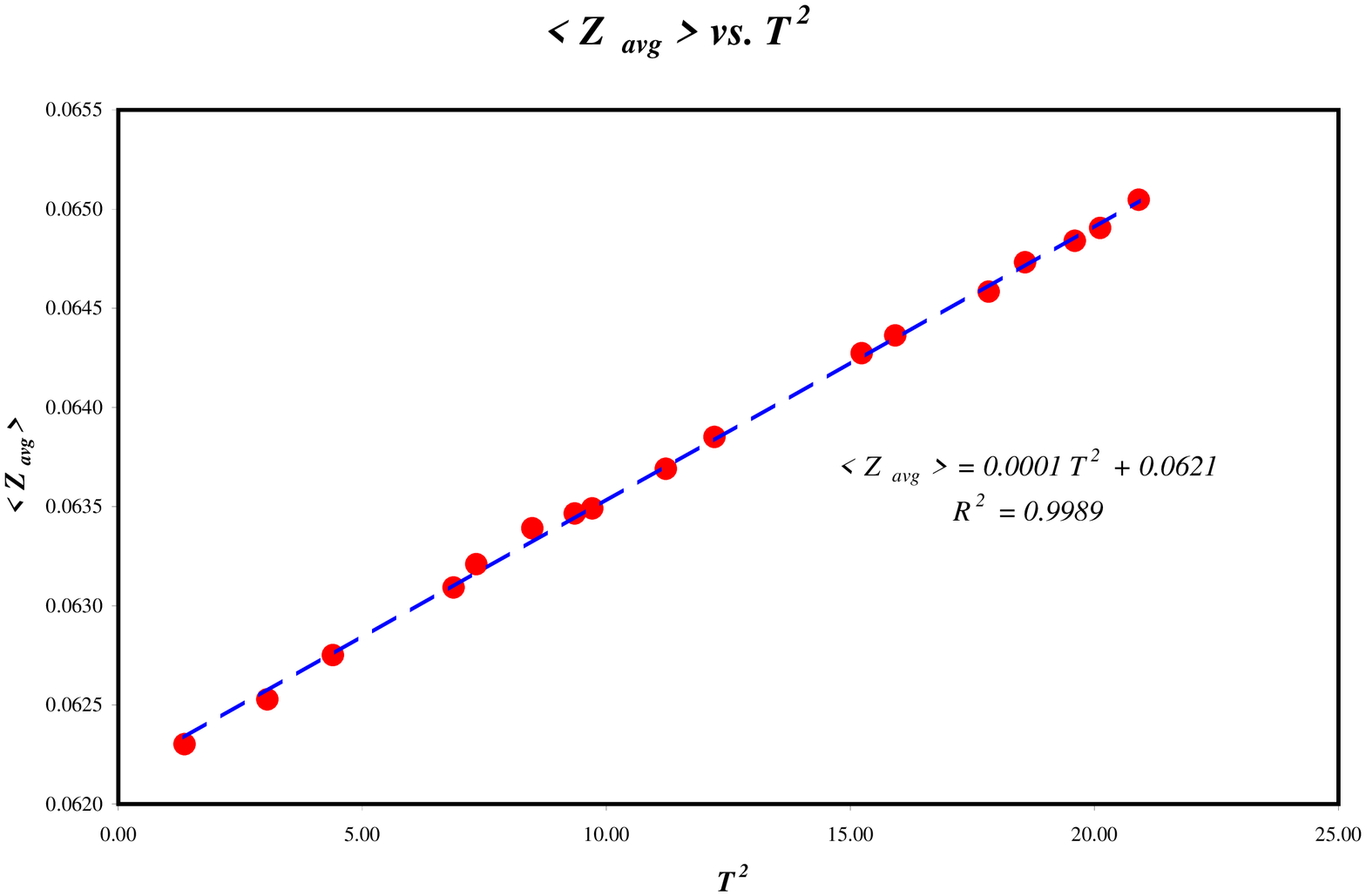,height=15cm}
\end{center}
\end{figure}

\newpage
\pagebreak

\begin{figure}
\begin{center}
\epsfig{file=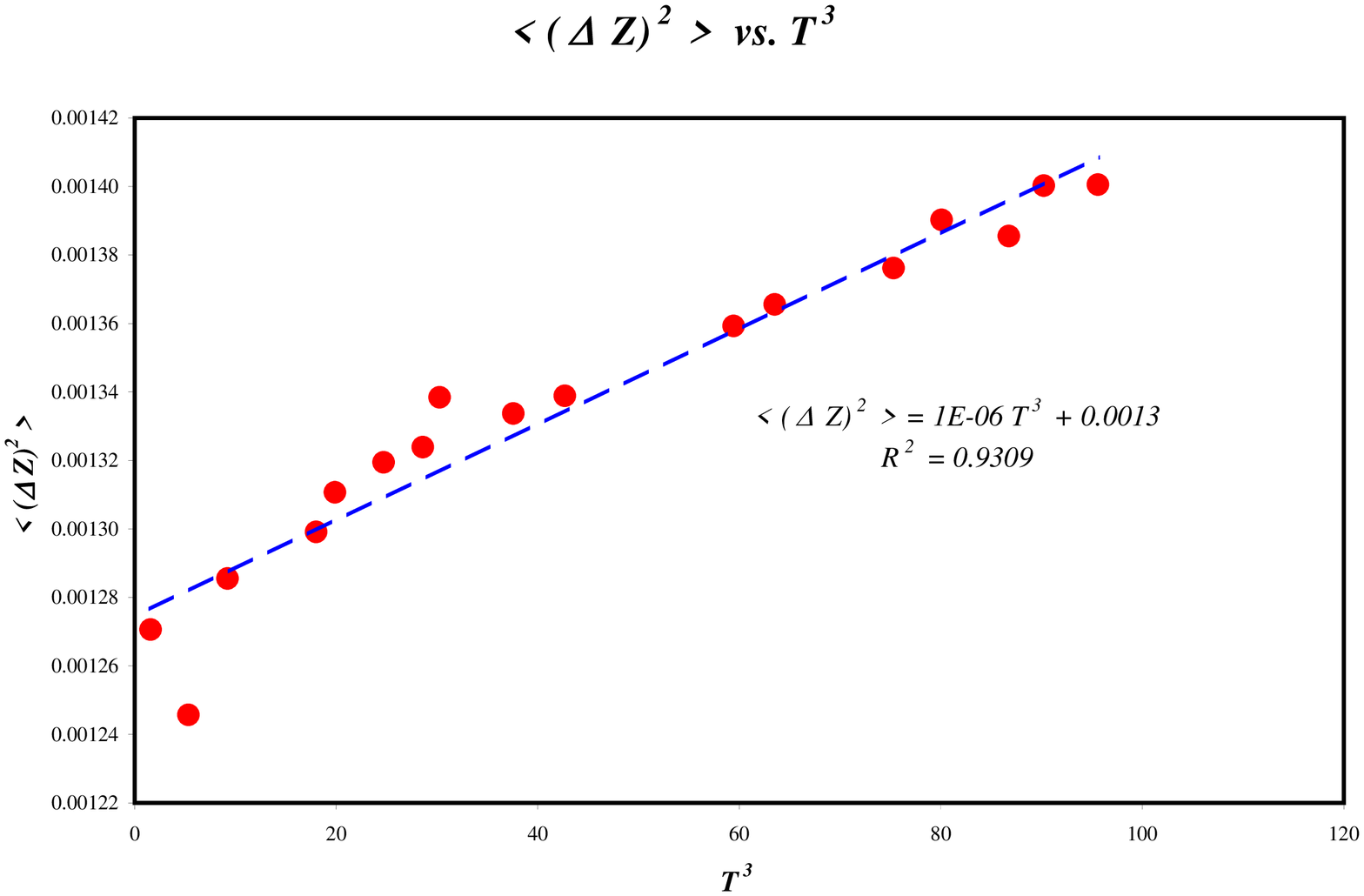,height=15cm}
\end{center}
\end{figure}

\end{document}